\documentclass[12pt]{article}

\usepackage{amssymb,amsmath,mathrsfs,subfigure,epsfig,epstopdf}
\usepackage{amsmath}

\usepackage[latin1]{inputenc}

\usepackage{hyperref}
\usepackage{xcolor}
\usepackage{ulem}

\usepackage{booktabs}
\usepackage[latin1]{inputenc}

\newcommand{\be}{\begin{equation}}
\newcommand{\en}{\end{equation}}

\newcommand{\ii}{\textrm{i}}

\renewcommand{\vec}[1]{\boldsymbol{#1}}

\begin{document}
\setlength{\fboxsep}{1pt}


\title{Generalization of the Zabolotskaya equation to all incompressible isotropic elastic solids\footnote{\noindent We dedicate this paper to the memory of Peter Chadwick FRS, a true pioneer in the study of elastic wave propagation.}
}

\author{
Michel Destrade$^1$, Edvige Pucci$^2$, Giuseppe Saccomandi$^{2,1}$
\\[12pt]
$^1$School of Mathematics, Statistics and Applied Mathematics,  \\
NUI Galway, University Road Galway, Ireland; \\[0.2cm]
$^2$Dipartimento di Ingegneria, \\
Universit\`{a} degli studi di Perugia, 06125 Perugia, Italy. }

\date{}

 \maketitle

\noindent
\textbf{keywords:} non-linear elasticity, non-linear waves, harmonics, Za\-bo\-lot\-ska\-ya equation, multiple scales.

 



\begin{abstract}

We study elastic shear waves of small but finite amplitude, composed of an anti-plane shear motion and a general in-plane motion. 
We use a multiple scales expansion to derive an asymptotic system of coupled nonlinear equations describing their propagation in all isotropic incompressible non-linear elastic solids, generalizing the scalar Zabolotskaya equation of compressible nonlinear elasticity.  
We show that for a general isotropic incompressible solid, the coupling between anti-plane and in-plane motions cannot be undone and thus conclude that linear polarization is impossible for general nonlinear two-dimensional shear waves.
We then use the equations to study the evolution of a nonlinear Gaussian beam in a soft solid: we show that a pure (linearly polarised) shear beam source generates only odd harmonics, but that introducing a slight in-plane noise in the source signal leads to a second harmonic, of the same magnitude as the fifth harmonic, a phenomenon recently observed experimentally.
Finally we present examples of some special shear motions with linear polarisation.

\end{abstract}

\maketitle 


\section{Introduction}


Anti-plane shear motions are some of the simplest motions around to investigate the nonlinear equations of elastodynamics \cite{Horgan}.  
This framework is quite general to start with, but its main limitation becomes apparent as soon as we write down the equations of motion, because it turns out that there are very few materials that may sustain a pure state of anti-plane shear in the absence of body forces. 
In general the equations of elastodynamics reduce to an overdetermined system of partial differential equations. 
As summarized by Pucci and Saccomandi \cite{Pucci1, Pucci2} and Saccomandi \cite{Sacco}, this compatibility problem can only be resolved under some special circumstances, for certain classes of materials.

In nonlinear acoustics, the evolution of an initial disturbance with a well-defined direction of propagation has been studied (among others) for longitudinal waves by Zabolotskaya \cite{ZT} in compressible isotropic nonlinear elasticity and by Norris and Kostek \cite{Norris} in compressible anisotropic nonlinear elasticity (see the review by Norris \cite{Norris98}). 

For \textit{transverse} waves  in compressible isotropic nonlinear elasticity, the situation is more complex because transverse beams are undistorted in the second-order approximation. 
As shown by  Zabolotskaya \cite{ZT}, the strain energy has to be expanded up to the fourth order to obtain a nonlinear motion. 
The resulting scalar Zabolotskaya equation (Z-equation) describing  the propagation of small-but-finite amplitude  shear waves in nonlinear elasticity \cite{Cramer, Z} is now a standard \textit{model equation} \cite{Hamilton}. 

In nonlinear acoustics, there is a widespread belief that the Z-equation is a model equation valid for \textit{any} constitutive equation, although \textit{it is not}, simply because the Z-equation is based on anti-plane shear motion which, as we recalled above, is sustainable only by some restricted classes of theoretical models for solids. 
Indeed, Destrade et al. \cite{Destrade} showed that for nonlinear isotropic \textit{incompressible} solids, the scalar Z-equation is valid only in the so-called `generalized neo-Hookean solids' and for some other special subclasses of constitutive equations.
The strain energy of generalized neo-Hookean solids is restrictive and does not reflect real solids, because it depends on one strain invariant only, instead of the two required for general nonlinear isotropic incompressible solids (see also Horgan and Saccomandi  \cite{Horgan1}).

Hence, nonlinear shear waves with linear polarization do not exist in general isotropic incompressible solids, only in solids with a very special form of strain energy density which might not be representative of any real-world material.
In general, nonlinear shear waves necessarily couple in-plane to anti-plane motion.

Here we derive {\color{black} a generalization of the Z-equation} which works for any isotropic incompressible elastic solid (Section \ref{section2}).
As expected, \color{black} the resulting system of equations \color{black} couples an in-plane motion to the anti-plane shear motion \cite{Pusac}.
We use a multiple scale expansion to derive the system of equations, placing ourselves in the general framework of exact nonlinear hyperelasticity. 
We then make the link with weakly nonlinear elasticity, and recover the system of equations established by Wochner  et al. \cite{Woch08}, albeit in a different form, with two equations instead of three, involving a `stream' function, from which the in-plane motion and the transverse component of the motion can be deduced.
We also peruse the literature to demonstrate that so far, no real-world incompressible solid has been found such that in-plane motion is decoupled from out-of-plane motion \color{black} in general.
Note that such a decoupling is possible for some special motions, as we show at the end of the paper, see also \cite{Horgan1, Pucci1} for a detailed discussion on this point. \color{black}

In Section \ref{Gaussian Beams}, we analyse the data recently provided by {\color{black} Esp\'indola et al.} \cite{Espi} in their investigation of nonlinear waves propagating in gelatine. 
They showed experimentally that an initially  {\color{black}linearly polarized} transverse Gaussian beam generated odd harmonics, but they did not remark on the small peak observed at the second harmonic. 
Here we show that it can be explained by considering that the initial Gaussian beam  possesses a small amount of {\color{black} in-plane \textit{noise}, in a sense clarified in Section \ref{Gaussian Beams}},  of one order of magnitude smaller than the out-of-plane component. 

Finally in Section \ref{Section4} we present  examples of \color{black} anti-plane \color{black} linear polarisation that can be achieved for some special nonlinear motions.


\section{The generalized Zabolotskaya (GZ) system}  
\label{section2}



\subsection{Derivation in exact nonlinear elasticity}


We denote by $\vec{F}=\partial \vec{x}/\partial \vec{X}$  the deformation gradient associated with the motion $\vec{x}=\vec{x}(\vec{X},t)$, where $\vec X$ and $\vec x$ are the positions in the reference and current configurations, respectively. 
The left Cauchy-Green strain tensor is $\vec{B}=\vec{FF}^T$ \color{black} and the first two principal strain invariants are $I_1 = \text{tr}\vec{B}$ and $I_2= [(\text{tr}\vec{B})^2 - \text{tr}(\vec B^2)]/2$. \color{black} 

In \textit{all generality}, the strain energy density $W$ \color{black} for isotropic incompressible hyperelastic materials  is a function of those \color{black} two invariants only:  $W=W(I_1, I_2)$, say.
Then the Cauchy stress tensor $\vec{T}$ is \cite{Rivlin}
\be \label{1}
\vec{T}=-p \vec{I} + 2  \left(\partial W / \partial I_1\right) \vec{B} - 2  \left(\partial W / \partial I_2\right) \vec{B}^{-1},
\en
where $p$ is a Lagrange multiplier associated with the incompressibility constraint, which reads locally as $\text{det}(\vec{F}) = 1$. 

The equations of elastodynamics in the absence of body forces read
\be \label{3}
\text{div }\vec{T} = \rho \, \vec{a},
\en
where $\rho$ is the (constant) mass density and $\vec{a}$ the acceleration vector. 
Equivalently, in terms of the nominal stress tensor $\vec{P} \equiv \vec{TF}^{-T}$, we can write them as
\be \label{4}
\text{Div }\vec{P} = \rho \, \partial^2 \vec{x}/\partial^2 t.
\en

In this paper we take $X$ as the direction of propagation and consider the following class of two-dimensional motions
\begin{equation} \label{5} 
x=X+\tilde{u}(X,Y,t), \qquad
y=Y+\tilde{v}(X,Y,t), \qquad
z=Z+\tilde{w}(X,Y,t),
\end{equation}
describing an anti-plane shear motion $\tilde{w}(X,Y,t)$ superimposed onto an in-plane motion with components $\tilde{u}(X,Y,t)$ and $\tilde{v}(X,Y,t)$.

For these motions we compute the components of the deformation gradient as
\be
\vec F = \begin{bmatrix}
1+\tilde u,_{X} & \tilde u,_{Y} & 0 \\
\tilde v,_{X} & 1+\tilde v,_{Y} & 0 \\
\tilde w,_{X} & \tilde w,_{Y} & 1
\end{bmatrix},
\en
in the $\vec e_i \otimes \vec E_j$ basis, where ($\vec E_1, \vec E_2, \vec E_3$) and ($\vec e_1, \vec e_2, \vec e_3$) are the unit vectors along the Cartesians axes ($X,Y,Z$) and ($x,y,z$), respectively, and the comma denotes partial differentiation.
From this expression we deduce
\begin{align}
& I_1 = (1+\tilde u,_{X})^2 + \tilde u,_{Y}^2 + \tilde v,_{X}^2 + (1+ \tilde v,_{Y})^2 + 1 + \tilde w,_{X}^2 + \tilde w,_{Y}^2, \notag \\
& I_2 = I_1 \, + (\tilde u,_{X} \tilde w,_{Y} - \tilde u,_{Y} \tilde w,_{X} {\color{black}+}  \tilde w,_{Y})^2 \notag\\
& \qquad \quad \, +  (\tilde v,_{X} \tilde w,_{Y} - \tilde v,_{Y} \tilde w,_{X} - \tilde w,_{X})^2 -  \tilde w,_{X}^2 - \tilde w,_{Y}^2 {\color{black},}
\end{align}
\color{black} and the incompressibility condition $\det \vec F=1$ reads \color{black}
\be \label{iso}
\tilde{u},_{X}+\tilde{v},_{Y} + \tilde{u},_{X}\tilde{v},_{Y}-\tilde{u},_{Y}\tilde{v},_{X} = 0.
\en
Note that we used this condition in computing $I_2$ above.

Our goal is to derive an asymptotic system able to describe two-dimensional shear waves. 
To this end we introduce a small parameter $\epsilon$ such that the amplitudes can be written as 
\be \label{amplitude}
  \tilde{u}=\epsilon u, \qquad \tilde{v}=\epsilon v, \qquad \tilde{w}=\epsilon w,
  \en
where $u$, $v$, $w$ are functions of $X,Y,t$ only, and are of order zero. 
Then we introduce the following scalings 
\be \label{scalings}
\chi=\epsilon^2 c X, \qquad \eta=\epsilon  Y,  \qquad \tau=t -  X/c,  \qquad
\tilde{p} = \epsilon^2 p,
\en
where  $c$ is the  speed of linear (infinitesimal) transverse elastic waves.

We will conduct asymptotic expansions up to $\mathcal{O}(\epsilon^3)$ and neglect terms of higher orders.  
We point out that this expansion procedure is the usual expansion of nonlinear acoustics,  see Norris \cite{Norris}, instead of the slow time scaling $t'=\epsilon^2 t$, which is often found in Continuum Mechanics \cite{Nariboli}. 
\color{black}These choices are equivalent, as they correspond to mapping the \textit{initial} conditions into a \textit{source} term and vice-versa.
\color{black}

It is now convenient  to introduce a ``stream function'' $\psi = \psi(X,Y,t)$ such that  \cite{Pusac} 
\be \label{ebis}
{u} =c \, \psi,_{Y}, \qquad {v} = -c \, \psi,_{X}. 
\en
Then, using the chain rule we obtain
\be \label{etris}
\tilde{u}=\epsilon^2  c\, \psi,_\eta, \qquad \tilde{v}=  \epsilon \psi,_\tau - \epsilon^3 c^2 \psi,_\chi,
\en
and the incompressibility equation \eqref{iso} is automatically satisfied at order $\mathcal{O}(\epsilon^3)$.  
Notice how the scalings \eqref{amplitude}-\eqref{scalings} ensure that the shear motion is dominant and the in-plane motion is small in comparison, with amplitudes at least one order of magnitude smaller, see \eqref{etris}. 

Further, we find the following expansions for the principal invariants,  
\be
I_1 = I_2 = 3 + \epsilon^2 J, \qquad \text{where} \qquad J = [(\psi,_{\tau \tau})^2 + (w,_\tau)^2]/c^2,
\en
is a non-dimensional quantity.
We can then expand the derivatives of the strain energy density as 
\be
  \partial W / \partial I_1 = \gamma_0 + \epsilon^2 \gamma_1 J, 
\qquad  \partial W / \partial I_2 = \lambda_0 + \epsilon^2 \lambda_1 J,
\en
where the constants $\gamma_0$, $\gamma_1$, and $\lambda_0$, $\lambda_1$ are defined as follows, 
\begin{align}
& \gamma_0 =  \left. \dfrac{\partial W}{\partial I_1} \right|_{I_1=I_2=3}, && 
\gamma_1 = \left. \left(\dfrac{\partial^2 W}{\partial I_1^2} + \dfrac{\partial^2 W}{\partial I_1 \partial I_2}\right) \right|_{I_1=I_2=3},
\notag \\[6pt]
& \lambda_0 =  \left. \dfrac{\partial W}{\partial I_2} \right|_{I_1=I_2=3}, && 
\lambda_1 = \left. \left(\dfrac{\partial^2 W}{\partial I_2^2} + \dfrac{\partial^2 W}{\partial I_1 \partial I_2}\right) \right|_{I_1=I_2=3}.
\end{align}

Enforcing continuity with linear isotropic incompressible elasticity, we find that \cite{Rivlin}
\begin{equation}
\gamma_0 + \lambda_0 = \mu/2,
\end{equation}
where $\mu = \rho c^2$ is the  {\color{black} infinitesimal} shear modulus (the second Lam\'e coefficient).

Now we introduce the following non-dimensional coefficients $\beta_2$ and $\beta_3$,
\begin{align}
& \beta_2 =  -\frac{\lambda_0}{\gamma_0+\lambda_0} 
 = -  \dfrac{2}{\mu}\left. \dfrac{\partial W}{\partial I_2} \right|_{I_1=I_2=3}, 
 \notag \\[10pt]
&
\beta_3 = \frac{3}{2} \, \frac{\gamma_1+\lambda_1}{\gamma_0 + \lambda_0} 
= \dfrac{3}{\mu} \left. \left(\dfrac{\partial^2 W}{\partial I_1^2} + 2 \dfrac{\partial^2 W}{\partial I_1 \partial I_2} + \dfrac{\partial^2 W}{\partial I_2^2} \right) \right|_{I_1=I_2=3}.
\end{align}
We then find (details not reproduced here) that to order $\epsilon^3$, the equations of motion read 
\begin{align}   \notag
 & \psi,_{\tau \tau \eta \eta} + 2 \psi,_{\tau \tau \tau \chi} + \dfrac{2\beta_3}{3c^2} (J \psi,_{\tau \tau}),_{\tau \tau}  - \dfrac{\beta_2}{c^2} (w,_{\tau \tau \tau} w,_\eta - w,_\tau w,_{\eta \tau \tau}) = 0, \notag
  \\[6pt]
  & w,_{\eta \eta} + 2 w,_{\tau \chi} + \dfrac{2\beta_3}{3c^2} (J w,_{\tau}),_\tau \notag \\
 & \qquad - \dfrac{\beta_2}{c^2} [\psi,_{\tau \tau \tau} w,_\eta-\psi,_{\eta \tau \tau} w,_\tau+2(\psi,_{\tau \tau} w,_{\eta \tau}-\psi,_{\eta \tau} w,_{\tau \tau})]=0.
\label{eq3}  \end{align} 

This is the \textit{Generalised Zabolotzkaya system} (GZ system), describing transverse waves travelling in any incompressible isotropic solid.
Once a solid is specified by a given strain energy density $W$, the constants $\beta_2$ and $\beta_3$ are computed from the formulas above, and the GZ equations form a system of two coupled nonlinear partial differential equations for $\psi$ and $w$. 
Once solved, it yields the displacement components $u$, $v$ (from \eqref{ebis}) and $w$ and the motion is described in its entirety.

It was first established by Wochner et al. \cite{Woch08} in the context of weakly nonlinear elasticity, with which we now connect.


\subsection{Connection with weakly nonlinear elasticity}


In all generality, we may expand the strain energy density $W$ in a Rivlin series, as \cite{Rivlin}
\be
W = \sum_{i+j=1}^\infty C_{ij}(I_1-3)^i(I_2-3)^j,
\en
where the $C_{ij}$ are constants. 
We then find that 
\be
\beta_2 = -\dfrac{2}{\mu}C_{01}, 
\qquad
\beta_3 = \dfrac{6}{\mu} \left(C_{20} + C_{11} + C_{02}\right).
\en

Now, at the same level of approximation, the $i+j=2$ Rivlin expansion of $W$ is equivalent \cite{DeGM10} to the following fourth-order Landau expansion of weakly-nonlinear elasticity \cite{Zabo04, DeOg10},
\be
W = \mu \: \text{tr}\left(\vec E^2\right) + \dfrac{A}{3}\: \text{tr}\left(\vec E^3\right) + D \: \text{tr}\left(\vec E^2\right)^2,
\en
where $\vec E = (\vec F^T \vec F - \vec I)/2$ is the Green-Lagrange strain tensor, and $A$, $D$ are the third- and fourth-order nonlinear Landau constants, respectively.
These constants are linearly connected \cite{DeGM10} to the $C_{ij}$, and eventually we find that 
\be
\beta_2 = 1 + \dfrac{A}{4\mu}, 
\qquad
\beta_3 =   \dfrac{3}{2} \left(1 + \dfrac{A/2+D}{\mu}\right).
\en
Hence $\beta_2$ invokes third-order elasticity only, and $\beta_3$, fourth-order elasticity.
With respect to stress-strain relationships, $\beta_2$ invokes quadratic nonlinearities, and $\beta_3$, cubic nonlinearities.
These constants were introduced in papers by Zabolotskaya and collaborators  \cite{Woch08,Spratt14, Spratt15}.
With this connection it is a simple matter to identify our formulation \eqref{eq3} of the equations of motion with that of Wochner et al. \cite{Woch08}.

If we were to study \textit{one-dimensional} plane shear waves (depending on only one space variable), then the derivatives with respect to $\eta$ would vanish from the GZ system \eqref{eq3} and the coefficient $\beta_2$ would play no role in the motion: there would be no quadratic nonlinearity for the wave and we would then recover the result of Zabolotskaya et al. \cite{Zabo04}, with the same coefficient of cubic nonlinearity $\beta_3$.

Here we are dealing with \textit{two-dimensional} shear waves, and the system shows a strong coupling between the three components of the wave \color{black} in general\color{black}.
Mathematically speaking, there are several ways to simplify the Generalised Z-system (GZ-system) of Equation \eqref{eq3} into decoupled equations, by playing on, and taking special values of the constants $\beta_2$ and $\beta_3$.
For instance \cite{Woch08, Destrade} by taking $\beta_2 = 0$: in that special case, the GZ-system decouples into an equation for $\psi$ and an equation for $w$. 
But solids with  the special property $\beta_2=0$ do not exist in the real world. 
We show this in Table \ref{table1}, where we computed the constants $\beta_2$ and $\beta_3$ from several experimental sources.
The conclusion is that in general isotropic incompressible solids, \textit{quadratic nonlinearities cannot be separated from cubic nonlinearities} when it comes to two-dimensional shear wave motion, contrary to what is postulated in the original paper by Wochner et al. \cite{Woch08}, and pursued by several works that followed, see for example \cite{Woch08b, Pinto10, Giam16,Pinton, Wang17}. 

However, for some \textit{special motions}, linear or plane polarisation in the transverse plane can be decoupled from the longitudinal motion: we present such examples in Section \ref{Section4}.
But first we study the effects of the coupling in the GZ-system on the evolution of a shear Gaussian beam and the generation of higher-order harmonics.

\begin{table}[!htbp]
\centering
\begin{tabular}{l | l c c l }
{Reference}   & {Material} &  $\beta_2 = 1 + \frac{A}{4\mu}$ & $\beta_3=   \frac{3}{2} \left(1 + \frac{A/2+D}{\mu}\right)$   \\
\bottomrule
\toprule

Catheline et al.  		&  phantom gel 1  & $-0.78 \pm 0.14$ 	&   \\
(2003) \cite{Catheline03}	&  phantom gel 2  & $-2.98 \pm 0.19 $ 	&   \\
  					&  phantom gel 3  & $-0.76 \pm 0.03 $ 	&   \\
\bottomrule
\toprule
R\'enier et al. 			&  5\% gelatine gel  & $-0.43 \pm 0.07$ &  $4.03 \pm 0.66$  \\
(2008) \cite{Renier08} 	&  7\% gelatine gel  & $0.33 \pm 0.01$ &   $2.50 \pm 1.28$ \\
\bottomrule \toprule
Latorre et al.  			&  soft phantom gel 1  & $-3.07 \pm 0.10$  	&   \\
(2012)	\cite{Latorre12}	&  soft phantom gel 2  & $0.44 \pm 0.07$ 	&   \\
					&  soft phantom gel 3  & $0.50 \pm 0.05$ 	&   \\
					&  beef liver 1  & $-9.03 \pm 2.68$ 		&   \\
					&  beef liver 2  & $-14.45 \pm 5.47$ 		&   \\
					&  beef liver 3  & $-17.02 \pm 4.57$ 		&   \\
\bottomrule \toprule
Jiang et al.  			&  pig brain (left)  	& $-0.43 \pm 0.11$  	&   $1.41 \pm 0.02$ \\
(2015)	\cite{jiang15}	&  pig brain (right)	& $-0.46 \pm 2.01$ 	&   $0.46 \pm 0.07$ \\
\bottomrule
\end{tabular}
\caption{
\small
Coefficients of quadratic  and cubic  nonlinearity for two-dimensional shear wave motion in some soft incompressible solids. The values are calculated from elastic wave measurements of $\mu$ (initial shear modulus) and $A$, $D$ (Landau coefficients). 
}
\label{table1}
\end{table}


\section{Gaussian beams in soft incompressible solids}
\label{Gaussian Beams}

\begin{figure}[h!]
\centering
\includegraphics[width=0.8\textwidth]{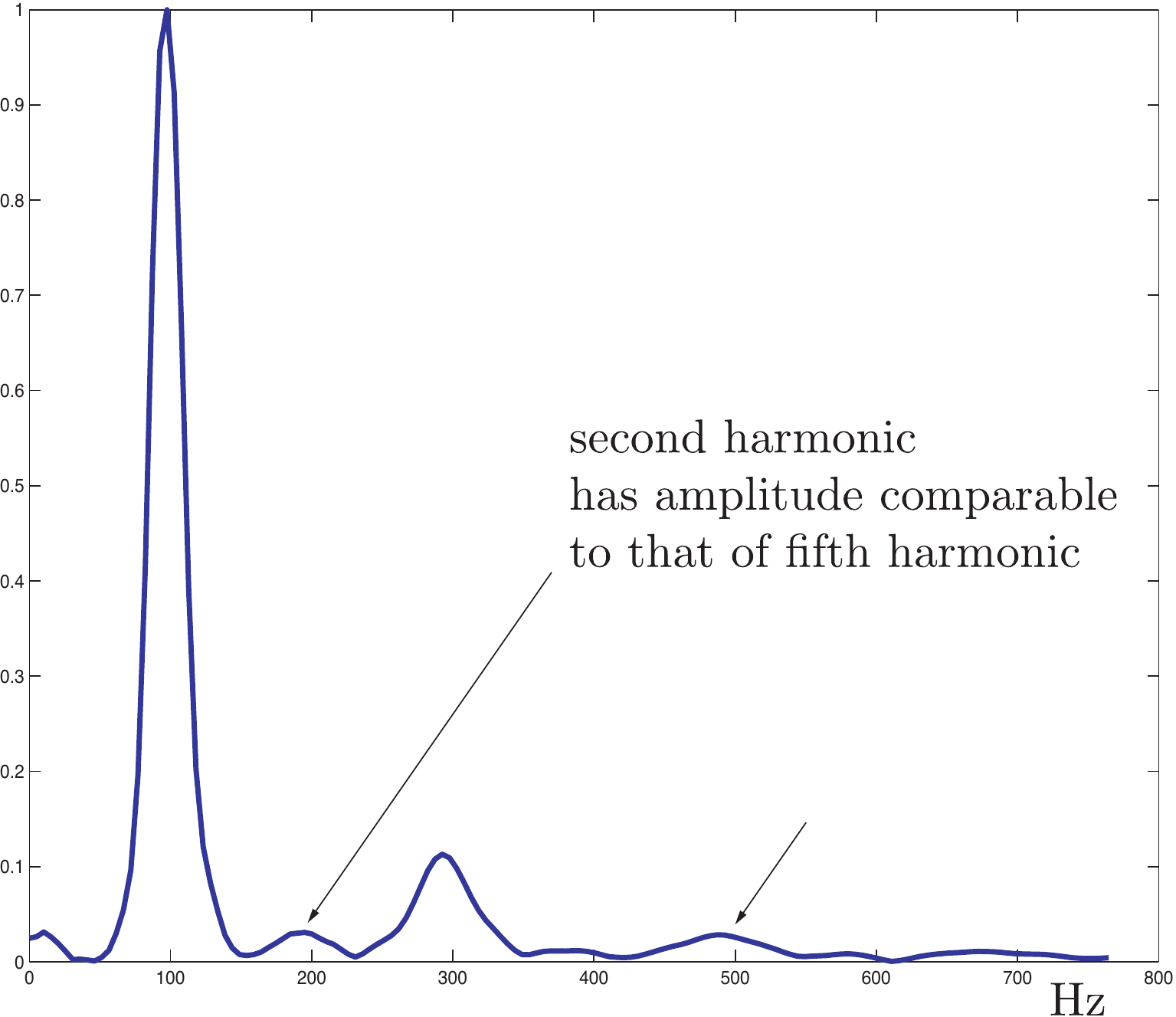}
\caption{
{\footnotesize
Spectrum generated by a large shear excitation in gelatine: experimental results of {\color{black} Esp\'indola et al.} \cite{Espi} (shared by G. Pinton and D. Esp\'indola).
}
}
\label{fig-pinton}
\end{figure}

  
{\color{black} Esp\'indola et al. \cite{Espi}} recently generated and measured nonlinear shear waves in gelatine. 
Figure \ref{fig-pinton} reproduces their data in the case of a high excitation amplitude generating a typical cubically nonlinear shock profile. 
The transmitted fundamental frequency at 100 Hz has the largest amplitude in the normalised spectrum, and some energy is generated at the (smaller) third harmonic (300 Hz) and at the (minor) fifth harmonic (500 Hz). 
Something that went un-noticed in the paper is that there is also a {\color{black} peak} at the second harmonic (200 Hz), of comparable amplitude to the one of the fifth harmonic at $500$ Hz. 


In this section we show that a shear wave beam source condition \color{black} which is \color{black} linearly polarised in the $Z-$direction (i.e. $\psi(0, \eta, \tau)\equiv 0$) does not produce a second harmonic in general, even when $\beta_2 \ne 0$ as is the case for real incompressible solids (in their modelling, {\color{black} Esp\'indola et al. \cite{Espi}}   take $\beta_2=0$).  
On the other hand, we show that if the polarisation of the shear wave beam is slightly {\color{black} misaligned, and therefore in our language \textit{`noisy'},} and allows for some small amplitude variations in the $XY-$plane (i.e. $\psi (0, \eta, \tau) \ne 0$), then a second harmonic is generated, with an amplitude comparable to that of the fifth harmonic.

We consider a regular perturbative solution of the GZ system \eqref{eq3}  via  a new small parameter $\varepsilon$, 
\begin{align} \notag
&w=\varepsilon w_1+ \varepsilon^{2} w_2+\varepsilon^{3} w_3+\varepsilon^{4} w_4+\varepsilon^{5} w_5+   \ldots, \\ \notag
& \psi= \varepsilon \psi_{1}  + \varepsilon^{2} \psi_{2}+\varepsilon^{3} \psi_{3}+\varepsilon^{4} \psi_{4}+\varepsilon^{5} \psi_{5}+ \ldots ,\notag
\end{align}
where $w_i$, $\psi_i$ are functions to be determined at each order.
For a general reference on this perturbation method, we refer to the books by Blackstock and Hamilton \cite{Hamilton} or by Naugolnykh and Ostrovski \cite{Ostro}.


\subsection{Pure {\color{black} anti-plane} shear beam}


First we take the source data (at $\chi=0$)  to be a \textit{pure anti-plane Gaussian beam}, i.e.
\begin{align} \label{per4} 
& w(0, \eta, \tau)= \varepsilon A_1 \exp(-\omega^2 \eta^2/c^2) \sin(\omega \tau), \notag \\
& \psi(0, \eta, \tau) \equiv 0,
\end{align}
where $A_1$ is a constant and $\omega$ is the frequency of the beam (or equivalently, $c/\omega$ is the effective source radius). 

By substitution in the GZ system \eqref{eq3} we obtain at the first order in $\varepsilon$,
\begin{equation} \label{per1}
w_{1, \eta \eta} + 2 w_{1,\tau \chi}=0, \qquad \psi_{1, \eta \eta \tau \tau } + 2 \psi_{1,\tau \tau \tau \chi}=0,
\end{equation}
two equations that are independent of the material parameters $\beta_2$ and $\beta_3$.
Clearly we may take $\psi_1 \equiv 0$ to satisfy the second equation, keeping a pure {\color{black} anti-plane} shear beam at first order.
We look for a solution to the first equation in the form
\begin{equation} \label{sol1}
w_1(\chi, \eta, \tau) = W_1(\chi, \eta)\exp(\ii \omega \tau)+ \text{c.c.},
\end{equation}
where $\ii = \sqrt -1$,  $W_1$ is a complex function, and `c.c.' stands for `complex conjugate'. 
Then \eqref{per1}$_1$ reduces to
\begin{equation} \label{parabolic}
W_{1, \eta \eta} + 2 \ii \omega W_{1, \chi}=0.
\end{equation}
The solution to this parabolic equation, subject to the condition \eqref{per4}, is  \cite{Ostro}
\begin{align} \label{W1}
 W_1=A_1H(\chi, \eta)\left[\sin Q(\chi, \eta)+\ii \cos Q(\chi, \eta) \right], 
 \end{align}
where $H$ and $Q$ are the following real functions
\begin{align} \label{pezzi} 
& H(\chi,\eta)= \dfrac{ \exp\left(-\dfrac{c^2 \eta^2}{{c^4}/{\omega^2}+4 \chi^2} \right)}{\left({c^4}/{\omega^2}+4 \chi^2\right)^{1/4}},
\notag \\[4pt]
& Q (\chi,\eta)= \frac{2 \chi \omega  \eta^2}{{c^4}/{\omega^2}+4 \chi^2}- \tfrac{1}{2} \arctan\left(2 \frac{\chi \omega}{c^2}\right).
\end{align}

Now we move on to order $\mathcal O  \left(\varepsilon^2\right)$. 
We notice that our first order solution $w_1(\xi, \eta, \tau)$ is such that $w_{1, \tau \tau}=-\omega^2 w_{1}$; it follows that 
\begin{equation} \label{cruciale}
w_{1,\tau \tau \tau} w_{1,\eta}-w_{1,\tau} w_{1,\eta \tau \tau}\equiv 0.
\end{equation}
Then the first equation of the GZ system \eqref{eq3} at that order reduces to
\begin{equation}
\psi_{2,\tau \tau \eta \eta} + 2 \psi_{2,\tau \tau \tau \chi} =0,
\end{equation}
for which we may  take $\psi_2 \equiv 0$ as a solution, maintaining the pure {\color{black} anti-plane} shear beam at the second order.
Then the second equation of the GZ system \eqref{eq3} at second order reduces to $w_{2,\eta\eta} + 2 w_{2,\tau \chi} = 0$, for which we may also take the trivial solution $w_2 \equiv 0$.

At order $\mathcal O  \left(\varepsilon^{3}\right)$ we obtain an equation for $w_3$ from the second  equation of the GZ system \eqref{eq3}, where the coefficient of cubic nonlinearity $\beta_3$ now plays a role:
\begin{equation} \label{per10}
w_{3,\eta \eta} + 2 w_{3, \tau \chi} + \dfrac{2 \beta_3}{3 c^2}  (w^3_{1, \tau})_\tau=0.
\end{equation}
Substituting our solution at order one, we have 
\begin{align} \label{per11} 
 w_{3,\eta \eta}+ 2 w_{3, \tau \chi}  = 2 \beta_3 (\omega^4/c^2) \left[W_1^2 \overline W_1 \exp(\ii \omega \tau) - W_1^3 \exp(3 \ii \omega \tau)  
+ \text{c.c.} \right],  
\end{align}
with $W_1$ given in \eqref{W1}.
Because of the linearity of this equation, we may look for solutions in the form 
\begin{equation} \label{sol3}
w_3 = W^{(1)}_3(\chi, \eta) \exp(\ii \omega \tau)+W^{(3)}_3(\chi, \eta) \exp(\ii 3 \omega \tau) + \text{c.c.},
\end{equation}
and obtain independent equations for the unknown first and third harmonic amplitude functions $W^{(1)}$ and $W^{(3)}$. 
For example, the determining equation for the first harmonic of the solution is the forced heat equation,
\begin{equation} \label{per12}
W^{(1)}_{3, \eta \eta} + 2 \ii \omega W^{(1)}_{3, \chi}= 2 \beta_3(\omega^4/c^2) A_1^2 H^2 W_{1}.  
\end{equation}
A similar forced heat equation determines $W^{(3)}_3$ and the solution is complete at order 3 for $w$. 
Here we see that, as is usual for transverse waves, the harmonic introduced by the initial condition is triplicated. 
At this same order we have the following equation for the in-plane components, coming from the first  equation of the GZ system \eqref{eq3}, 
\begin{equation}  \label{per2}   \psi_{3,\tau \tau \eta \eta} + 2 \psi_{3,\tau \tau \tau \chi}=0,
\end{equation} 
for which we may take $\psi_3\equiv0$, maintaining a pure {\color{black} anti-plane} shear beam at third order.

Moving on now to order $\mathcal O  \left(\varepsilon^4\right)$, we find for the first  equation of the GZ system \eqref{eq3},
 \begin{equation}  \label{per2bis} 
  \psi_{4,\tau \tau \eta \eta} + 2 \psi_{4,\tau \tau \tau \chi} 
 =\dfrac{\beta_2}{c^2} [w_{3,\tau \tau \tau} w_{1,\eta}-w_{3,\tau} w_{1,\eta \tau \tau}
  + w_{1,\tau \tau \tau} w_{3,\eta}-w_{1,\tau} w_{3,\eta \tau \tau}],
\end{equation}
and now the in-plane motion is excited (we checked by a trivial direct computation  that the coupling term in the brackets is not null.)
We point out that at this order ($\varepsilon^4$), the lowest order where the in-plane motion manifests itself, it is composed  of even harmonics only. 
Then the second equation of the GZ system \eqref{eq3} at fourth  order reduces to $w_{4,\eta\eta} + 2 w_{4,\tau \chi} = 0$, which we may solve by taking $w_4 \equiv 0$.

Here we see that the presence of the $\beta_2$ coefficient does eventually lead to  a coupling between in-plane and anti-plane wave components. 
The coupling is weak, in the sense that the anti-plane shear component $w$ of amplitude $\varepsilon$ is coupled to an in-plane function $\psi$ of magnitude $\varepsilon^4$.

The interaction between  the $\psi_4$ term of the in-plane motion and the anti-plane motion $w$ manifests itself at the next order, again thanks to the presence of the $\beta_2$ coupling term.  
Hence at order $ \mathcal O\left(\varepsilon^5\right)$ we find that the first equation of the GZ system \eqref{eq3} gives
\begin{align} \label{per20} 
& w_{5,\eta \eta} +  2  w_{5, \tau \chi} + \dfrac{2\beta_3}{c^2}(w_{1,\tau}^2w_{3,\tau}),_\tau
\\ \notag
& \qquad - \dfrac{\beta_2}{c^2} [\psi_{4,\tau \tau \tau} w,_{1,\eta}-\psi_{4, \eta \tau \tau} w_{1,\tau}+2(\psi_{4, \tau \tau} w_{1, \eta \tau}-\psi_{4, \eta \tau} w_{1, \tau \tau})]=0,
\end{align}
(and we checked that the bracketed term is not zero.) 
The same structure and sequences are repeated at higher orders.
Clearly, this means that only odd harmonics are generated: a pure (hypothetical) anti-plane Gaussian beam cannot generate even harmonics.

\begin{figure}[h!]
\centering
\includegraphics[width=0.44\textwidth]{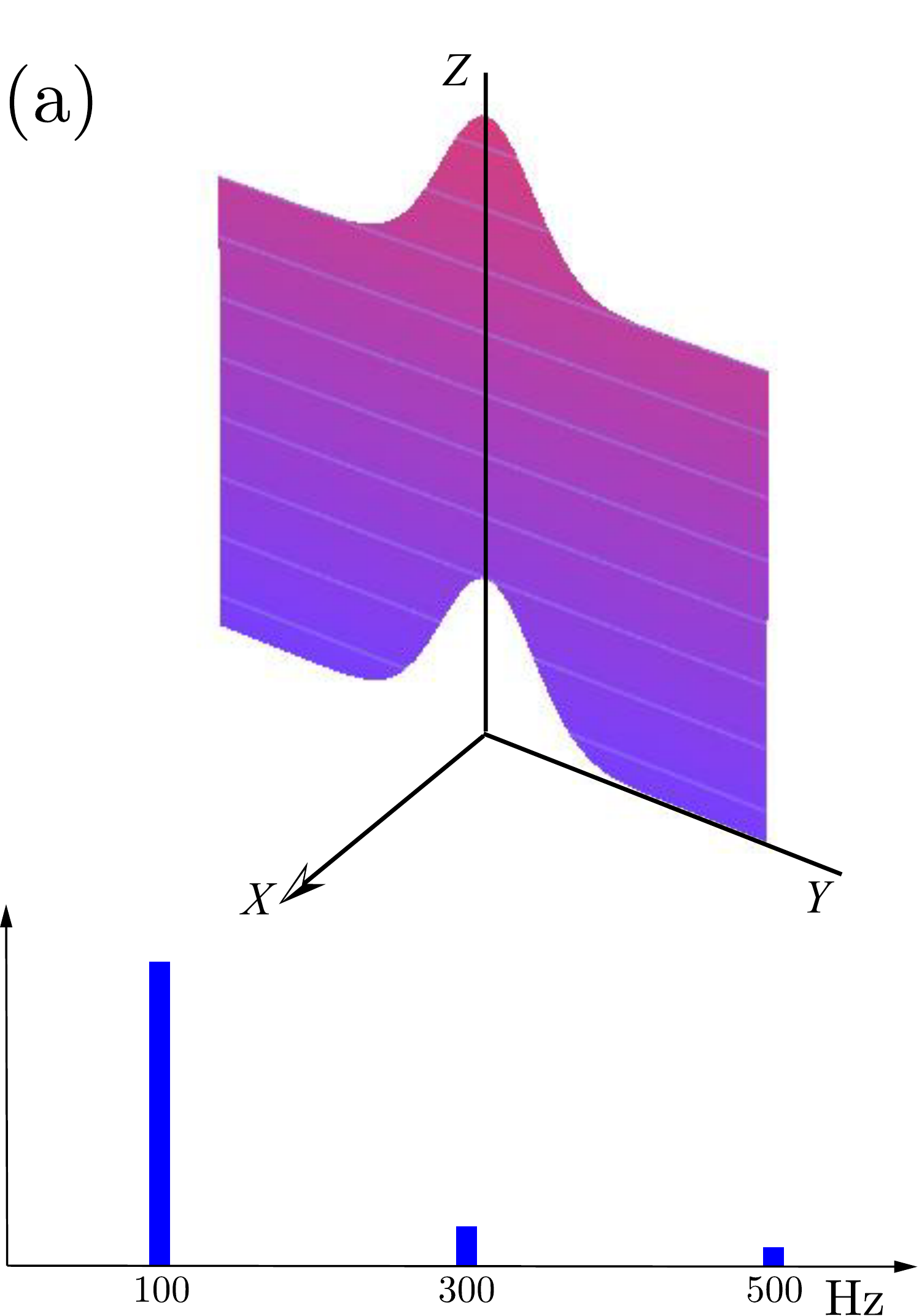} \qquad
\includegraphics[width=0.44\textwidth]{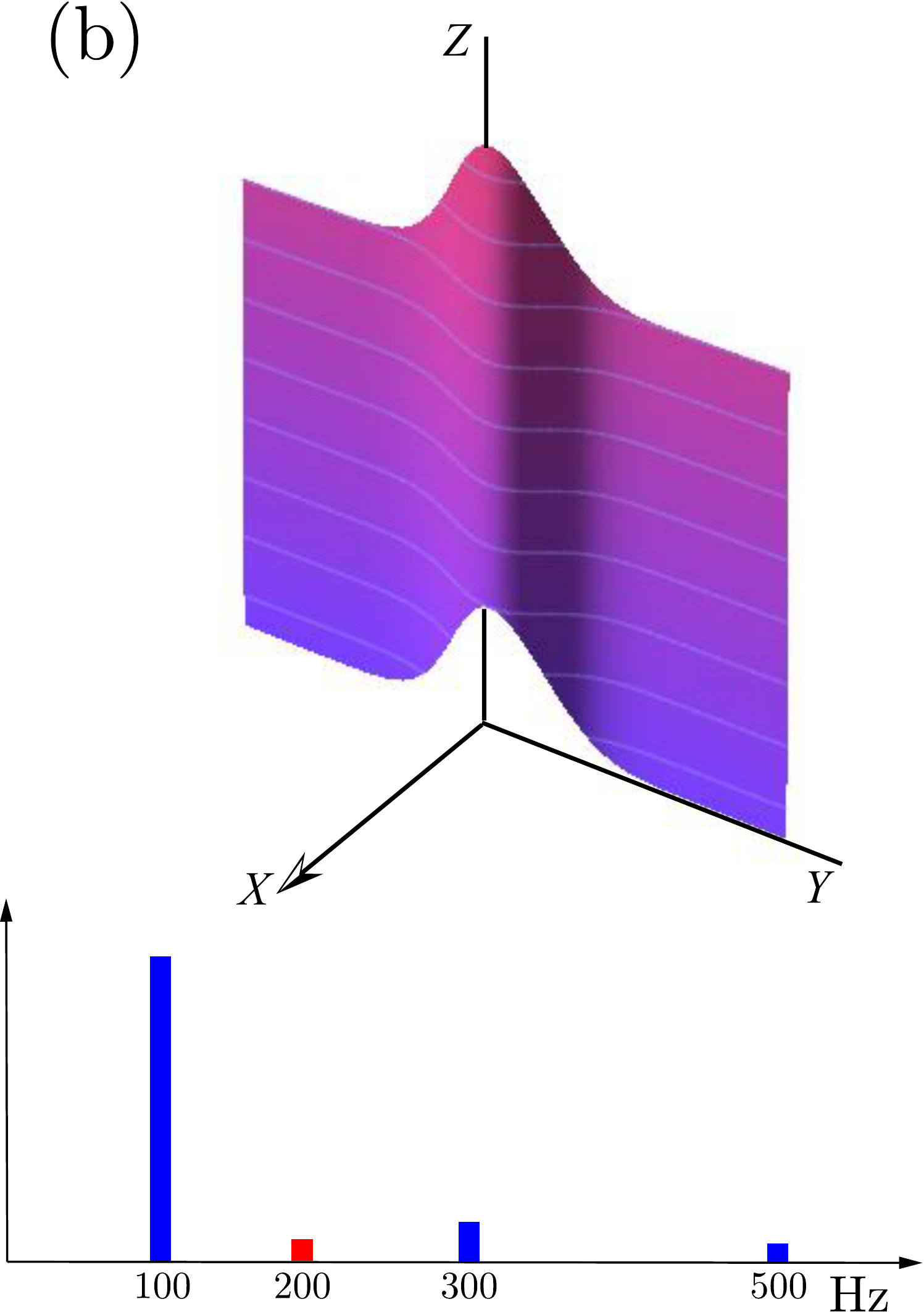}
\caption{
{\footnotesize
Sketches of the results in Section \ref{Gaussian Beams}:
(a) A pure anti-plane shear Gaussian beam generates only odd harmonics, at 100, 300, 500 Hz, while (b) An initially noisy  Gaussian beam (with a small in-plane component) additionally generates a harmonic at 200 Hz.
}
}
\label{fig-sketch}
\end{figure}


\subsection{Noisy shear beam}


Now let us consider the case  of a  source data which is \textit{not a pure {\color{black} anti-plane} shear Gaussian beam};
indeed,  it is impossible to achieve perfect beam focusing experimentally, and we expect a slightly noisy initial data, which we now model as
\begin{align} \label{per4bis}  
& w(0, \eta, \tau)= \varepsilon A_1 \exp(-\omega^2 \eta^2/c^2) \sin(\omega \tau), \notag \\
& \psi(0, \eta, \tau)= \varepsilon^2 B_2 \exp(-\omega^2 \eta^2/c^2) \sin(\omega \tau),
\end{align}
where $B_2$ is a constant.

In this case, in the first order of $\varepsilon$ we find the same solution as for the pure {\color{black} anti-plane} shear beam initial data, see previous section. 
At second order, the identity \eqref{cruciale} holds and the first equation of the GZ system \eqref{eq3} gives  the following equation for $\psi_2$,
\begin{equation}
\psi_{2, \eta \eta \tau \tau } + 2 \psi_{2, \tau \tau \tau \chi}=0,
\end{equation}
as in the previous section, but now with source value \eqref{per4bis}$_2$.
The solution is 
\begin{equation}
\psi_2 = B_2H(\chi, \eta)\left[\sin Q(\chi, \eta) + \ii \cos Q(\chi, \eta) \right] \exp(\ii \omega \tau) + \text{c.c.},
\end{equation}
where $H$ and $Q$ were defined in \eqref{pezzi}.
It follows by comparison with \eqref{sol1} and \eqref{W1} that $w_1$ and $\psi_2$ are proportional to each other:
\begin{equation} \label{crucialebi}
\psi_2 =(B_2/A_1)  w_1.
\end{equation} 
{\color{black} As before, $w_2=0$.}

At order $\mathcal (\varepsilon^3)$, the determining equation for $w_3$ reads 
\begin{multline} 
w_{3,\eta \eta} + 2 w_{3, \tau \chi} = - \dfrac{2\beta_3}{3c^2}    (w^3_{1, \tau})_{{\color{black},}\tau} \\ 
+ \dfrac{\beta_2}{c^2} [\psi_{2,\tau \tau \tau} w_{1,\eta}-\psi_{2, \eta \tau \tau} w_{1,\tau}+2(\psi_{2,\tau \tau} w_{1,\eta \tau}-\psi_{2, \eta \tau} w_{1, \tau \tau})]=0.
\end{multline}
Here it is an easy matter to check that  the bracketed term is identically zero. 
Therefore $w_3$ is given by the solution of \eqref{per10} obtained in the previous section.
Then the equation for $\psi_3$ is  \eqref{per2}, which we solve with $\psi_3 \equiv 0$.
 
 At  order $\mathcal{O}(\varepsilon^4)$, the first equation of the GZ system \eqref{eq3}  determines $\psi_4$ as the solution to 
\begin{multline}  \label{per21} 
  \psi_{4,\tau \tau \eta \eta} + 2 \psi_{4,\tau \tau \tau \chi} = - \dfrac{2\beta_3}{3c^2}   \left(\psi_{2, \tau \tau}w^2_{1, \tau} \right)_{{\color{black},}\tau \tau}  \\ 
  + \dfrac{\beta_2}{c^2}  [(w_{3,\tau \tau \tau} w_{1,\eta} - w_{3,\tau} w_{1,\eta \tau \tau})
  +(w_{1,\tau \tau \tau} w_{3,\eta}-w_{1,\tau} w_{3,\eta \tau \tau})].
\end{multline}
The  difference between this equation and the equation \eqref{per2bis} obtained at this order when the initial data is a pure anti-plane shear is the additional term proportional to $\beta_3$.
However, we see from Equation \eqref{crucialebi} that 
\begin{equation}
\left(\psi_{2, \tau \tau}w^2_{1, \tau} \right)_{{\color{black},}\tau \tau}
 = (B_2/A_1) \left(w_{1, \tau \tau}w^2_{1, \tau}\right)_{{\color{black},}\tau \tau} 
  =  [B_2/(3A_1)]\left(w^3_{1, \tau}\right)_{{\color{black},}\tau \tau \tau}.
\end{equation}
Now by a direct computation (not reproduced here) we find that the solution $\psi_4$ of \eqref{per21} contains the first, third and fourth harmonics.

Then the second equation of the GZ system \eqref{eq3} at the fourth order may be solved by taking $w_4 \equiv 0$, as in the previous section.

Now we write down the second equation of the GZ system \eqref{eq3}  at order $\mathcal O  \left(\varepsilon^5\right)$ for $w_5$. It reads
\begin{align} \label{per25} \notag
 w_{5,\eta \eta} +  2 w_{5, \tau \chi} =  &- \dfrac{2\beta_3}{\color{black}3\color{black}c^2} \left(w_{1,\tau}^2w_{3,\tau} \right){{\color{black},_\tau}} - \dfrac{2\beta_3}{3c^2}\left(\psi^2_{2, \tau \tau}w_{1, \tau} \right),_{\tau}
 \notag \\[6pt] 
 &  + \dfrac{\beta_2}{c^2} \left[\psi_{2,\tau \tau \tau} w_{3,\eta}-\psi_{\color{black}2\color{black}, \eta \tau \tau} w_{3,\tau}+2(\psi_{2, \tau \tau} w_{3, \eta \tau}-\psi_{2, \eta \tau} w_{3, \tau \tau})\right] 
 \notag
\\[6pt]
& + \dfrac{\beta_2}{c^2} [\psi_{4,\tau \tau \tau} w_{1,\eta}-\psi_{4, \eta \tau \tau} w_{1,\tau}+2(\psi_{4, \tau \tau} w_{1, \eta \tau}-\psi_{4, \eta \tau} w_{1, \tau \tau})].
\end{align}
Our goal here is to show that the solution $w_5$ of Equation \eqref{per25} contains the second harmonic. 
Clearly,  it  can only arise from the first bracketed term on the right-hand side, which is 
\begin{equation}
\psi_{2,\tau \tau \tau} w_{3,\eta}-\psi_{2, \eta \tau \tau} w_{3,\tau}+2(\psi_{2, \tau \tau} w_{3, \eta \tau}-\psi_{2, \eta \tau} w_{3, \tau \tau}).
\end{equation} 
Using \eqref{crucialebi}, we  see that this forcing term is proportional to
\begin{equation} \label{per26}
w_{1,\tau \tau \tau} w_{3,\eta}-w_{1, \eta \tau \tau} w_{3,\tau}+2(w_{1, \tau \tau} w_{3, \eta \tau}-w_{1, \eta \tau} w_{3, \tau \tau}).
\end{equation}
Further, taking the solution ${\color{black}W}_1$ in Equation \eqref{sol1} and the ${\color{black}W}^{(1)}_3$ component of \eqref{sol3}, we find the forcing term above to be 
\begin{align} \label{finale}
\exp(2\ii \omega \tau) \left[W_1W^{(1)}_{3,\eta}-W_{1,\eta}W^{(1)}_{3}\right] + \text{ c.c. }+\ldots,
\end{align}
where the ellipsis refers to higher harmonics terms.
Our claim is that 
\begin{equation}
W_1W^{(1)}_{3,\eta}-W_{1,\eta}W^{(1)}_{3}\neq 0.
\end{equation}
Indeed if the left-hand side above were zero, then integration would yield $W^{(1)}_3=\Gamma(\chi) W_1$ for some arbitrary function  $\Gamma$. 
Substitution into the forced heat equation \eqref{per12} and taking \eqref{parabolic} into account, we would then obtain
\begin{equation}
\Gamma'(\chi) = {\color{black}-} \ii \frac{\beta_{3}}{c^2}\omega^3 {\color{black}A_1^2}H^{\color{black}2\color{black}}(\chi, \eta),
\end{equation}
which is absurd in view of the dependence of  $H$ on $\eta$. 

The conclusion is that $w_5$ contains a second harmonic according to \eqref{per26}, in addition to the expected fifth harmonic coming out of the full solution. 
This modeling result is perfectly  aligned with the experimental data presented by {\color{black}Esp\'indola} et al. \cite{Espi}, see also Figure \ref{fig-pinton}.

\color{black}
We could of course go further in our asymptotic expansion. 
At higher orders, we would then expect the next even harmonics to be expressed, but with rapidly diminishing amplitudes, as can roughly be inferred from the experimental results in Figure \ref{fig-pinton}.
\color{black}


\section{Linear polarisation for special motions}
\label{Section4}


In the final section we explore some avenues to attain linear polarisation, simply by imposing that $\psi \equiv 0$ (and then $u=v=0$ by \eqref{ebis}, and only $w$ remains). 

With that assumption, the GZ-system  \eqref{eq3} reduces to the \textit{overdetermined system}, 
\be \label{s1}
w,_{\tau \tau \tau} w,_\eta-w,_\tau w,_{\eta \tau \tau}=0,  \qquad 
w,_{\eta \eta} + 2 w,_{\tau \chi} + \dfrac{2\beta_3}{3c^2}  (w^3{\color{black},}_{\tau}){\color{black},}_\tau=0. 
\en 
In general two equations cannot be solved simultaneously for a single unknown, but it might be possible that these two equations are compatible for \textit{special classes of solutions}. 

For example, consider the following ansatz,
\be \label{SC}
w=F(\chi, \zeta), \qquad \text{where} \quad \zeta= \tau+f(\chi, \eta),
\en
where $f$ is an arbitrary function.
Then the first equation in \eqref{s1} is the trivial identity and the system is no longer overdetermined. 
Introducing the ansatz in Equation \eqref{s1}$_2$, we obtain
\be \label{s1bis}
F,_{\zeta \zeta} (f,_\eta^2 +  2f,_\chi) + 2 F,_{\chi \zeta} + F,_\zeta f,_{\eta \eta} + \dfrac{2\beta_3}{3c^2}(F,_\zeta^3),_\zeta=0.
\en
Now we specialise the analysis to the cases where $f,_{\eta \eta}$ and $f,_\eta^2 + 2f{\color{black},}_\chi$ are functions of $\chi$ only, say    
\be  \label{s4}
f,_{\eta \eta}=f_1(\chi), \qquad f,_\eta^2 + 2f,_\chi=f_{\color{black}2\color{black}}(\chi),
\en
where $f_1$, $f_2$ are arbitrary functions.
In that case, it follows from \eqref{s4} that  
\be \label{s3}
f (\chi, \eta) =  \frac{1}{2 (\chi+c_0)} \eta^2+ \dfrac{c_1}{\chi+c_0} \eta+f_0(\chi),
\en
where $f_0$ is an arbitrary function and $c_0$,  $c_1$ are arbitrary constants.
The choice $c_0=c_1=0$ and $f_0 \equiv 0$ is remarkable, because it reduces Equation \eqref{s1bis}  to
\be
 \hat{F},_{\chi} + \frac{1}{2 \chi} \hat{F} +\frac{\beta_3}{3c^2} (\hat{F}^3),_{\zeta}=0,
\en
where $\hat{F}=F,_\zeta$. 
With the Sionoid-Cates transformation \cite{Sino},
\be
\tilde{\chi}=\ln \left(\frac{\chi}{c^2 \chi_0} \right), \qquad  
\hat{F}(\chi, \zeta)=\exp\left(-\tilde{\chi}/2\right) \tilde{F}(\tilde{\chi}, \zeta),
\en
where $\chi_0$ is a characteristic length, we rewrite the equation as an \textit{inviscid cubic Burger's equation},
\be
\tilde{F},_{\tilde{\chi}} + \frac{\beta_3 \chi_0}{c^2}\tilde{F}^2\tilde{F},_{\tilde\zeta}=0.
\en

The ansatz in \eqref{SC} determines a remarkable class of shear waves for which the overdetermined system \eqref{s1} admits solutions but this is not the only possibility. 
{\color{black}Indeed, } another possible class is given by `line solitary wave' solutions, where we take $w=W(k' \chi +k''\eta - \omega \tau)$. It is also interesting to point out that the first equation in \eqref{s1} is identically satisfied in the case
\be
w=\sum_{n=1}^{\infty} w_n(\xi, \eta) \exp(\ii n \omega t) + \text{c.c.}.
\en
Then the approximate solution corresponding to the source data  \eqref{per4} `uncouples'  the anti-plane and in-plane motions at any order.

\bigskip


\section*{Acknowledgements} 


EP and GS have been partially supported for this work by the Gruppo Nazionale per la Fisica Matematica (GNFM) of the Italian non-profit research institution Istituto Nazionale di Alta Matematica Francesco Severi (INdAM)  {\color{black} and the PRIN2017 project ``Mathematics of active materials: From mechanobiology to smart devices'' funded by the Italian Ministry of Education, Universities and Research (MIUR)}. We are most grateful to Gianmarco Pinton and David Esp\'indola for sharing the experimental data used to generate Figure \ref{fig-pinton}.


\begin{thebibliography}{99}
  
  
 
 
 

 
\bibitem{Cramer}
Cramer, M. S., \& Andrews, M. F. (2003). 
A modified Khokhlov-Zabolotskaya equation governing shear waves in a prestrained hyperelastic solid. 
The Journal of the Acoustical Society of America, 114, 1821-1832.


\bibitem{Catheline03}
Catheline, S., Gennisson, J.L. \& Fink, M. (2003). 
Measurement of elastic nonlinearity of soft solid with transient elastography. 
The Journal of the Acoustical Society of America, 114, 3087-3091.

\bibitem{DeGM10}
Destrade, M., Gilchrist, M. D., \& Murphy, J. G. (2010). 
Onset of non-linearity in the elastic bending of blocks. 
ASME Journal of Applied Mechanics, 
77, 061015. 
 
\bibitem{DeOg10}
Destrade, M., \& Ogden, R.W. (2010). 
On the third-and fourth-order constants of incompressible isotropic elasticity. 
Journal of the Acoustical Society of America, 
128, 3334-3343.
 
\bibitem{Destrade}
Destrade, M., Goriely, A., \& Saccomandi, G. (2010). Scalar evolution equations for shear waves in incompressible solids: a simple derivation of the Z, ZK, KZK and KP equations. Proceedings of the Royal Society of London A, 467, 1823-1834. 




\bibitem{Espi}
\color{black}
Esp\'indola, D., Lee, S. \& Pinton, G. (2017). 
Shear shock waves observed in the brain. Physical Review Applied, 8, 044024.
\color{black}

\bibitem{Giam16}
Giammarinaro, B., Coulouvrat, F. \& Pinton, G. (2016). 
Numerical simulation of focused shock shear waves in soft solids and a two-dimensional nonlinear homogeneous model of the brain. 
Journal of Biomechanical Engineering, 138, 041003.

\bibitem{Pinton}
Giammarinaro, B., Esp\'indola, D., Coulouvrat, F., \& Pinton, G. (2018). 
Focusing of shear shock waves. 
Physical Review Applied, 9, 014011.

\bibitem{Hamilton}
Hamilton, M. F., \& Blackstock, D. T. (Eds.). (1998). 
Nonlinear Acoustics (Vol. 427). San Diego: Academic press.


\bibitem{Horgan}
Horgan, C. O. (1995). 
Anti-plane shear deformations in linear and nonlinear solid mechanics. 
SIAM Review, 37, 53-81.

\bibitem{Horgan1}
Horgan, C. O., \& Saccomandi, G. (2003). 
Superposition of generalized plane strain on anti-plane shear deformations in isotropic incompressible hyperelastic materials. 
Journal of Elasticity, 73, 221-235.


 
\bibitem{jiang15}
Jiang, Y., Li, G., Qian, L.X., Liang, S., Destrade, M. \& Cao, Y. (2015). 
Measuring the linear and nonlinear elastic properties of brain tissue with shear waves and inverse analysis. 
Biomechanics and Modeling in Mechanobiology, 14, 1119-1128.

\bibitem{Latorre12}
Latorre-Ossa, H., Gennisson, J.L., De Brosses, E. \& Tanter, M. (2012). 
Quantitative imaging of nonlinear shear modulus by combining static elastography and shear wave elastography. 
IEEE Transactions on Ultrasonics, Ferroelectrics, and Frequency Control, 59, 833-839.



\bibitem{Ostro} 
Naugolnykh, K., \& Ostrovsky, L. (1998). Nonlinear wave processes in acoustics. Cambridge University Press. 


\bibitem{Nariboli}
Nariboli, G. A., \& Lin, W. C. (1973). A new type of Burgers' equation. ZAMM Zeitschrift f\"ur Angewandte Mathematik und Mechanik, 53, 505--510.
 
\bibitem{Norris}
Norris, A. N., \& Kostek, S. (1993). Nonlinear parabolic wave equations for solids. Advances in nonlinear acoustics: 13th ISNA, 463-471.

\bibitem{Norris98}
Norris, A. N. (1998). Finite amplitude waves in solids. In: \textit{Nonlinear Acoustics}, M.F. Hamilton, D.T. Blackstock (Eds.), Academic Press. 
 
\bibitem{Pinto10}
Pinton, G., Coulouvrat, F., Gennisson, J.L. \& Tanter, M., 2010. 
Nonlinear reflection of shock shear waves in soft elastic media. 
J. Acoust. Soc. Am.,  127, 683-691.
 
 \bibitem{Pucci1}
Pucci, E., \& Saccomandi, G. (2013). The anti-plane shear problem in nonlinear elasticity revisited. Journal of Elasticity, 113(2), 167-177.

\bibitem{Pucci2}
Pucci, E., Rajagopal, K. R., \& Saccomandi, G. (2015). On the determination of semi-inverse solutions of nonlinear Cauchy elasticity: The not so simple case of anti-plane shear. International Journal of Engineering Science, 88, 3-14.

\bibitem{Pusac}
Pucci, E., \& Saccomandi, G. (2018). A remarkable generalization of the Z equation, 
Mechanics Research Communications, to appear. 

\bibitem{Renier08}
R\'enier, M., Gennisson, J.L., Barri\`ere, C., Royer, D. \& Fink, M. (2008). 
Fourth-order shear elastic constant assessment in quasi-incompressible soft solids. 
Applied Physics Letters, 93, 101912.

\bibitem{Rivlin}
Rivlin, R.S., Barenblatt, G.I., \& Joseph, D.D. (1997). 
Collected papers of RS Rivlin (Vol. 1). Springer Science \& Business Media.
 
\bibitem{Sacco}
Saccomandi, G. (2016). DY Gao: Analytical solutions to general anti-plane shear problems in finite elasticity. Continuum Mechanics and Thermodynamics, 28(3), 915-918.


\bibitem{Sino}
Sionoid, P. N., \& Cates, A. T. (1994). The generalized Burgers and Zabolotskaya-Khokhlov equations: transformations, exact solutions and qualitative properties. In Proceedings of the Royal Society of London A: Mathematical, Physical and Engineering Sciences 447, 253--270.  

\bibitem{Spratt14}
Spratt, K.S. (2014). Second-harmonic generation and unique focusing effects in the propagation of shear wave beams with higher-order polarization (Doctoral dissertation, University of Texas).

\bibitem{Spratt15}
Spratt, K.S., Ilinskii, Y.A., Zabolotskaya, E.A. and Hamilton, M.F. (2015). 
Second-harmonic generation in shear wave beams with different polarizations. In AIP Conference Proceedings (Vol. 1685, No. 1, p. 080007). AIP Publishing.



\bibitem{Wang17}
Achenbach, J.D., Wang, Y. (2018). 
Far-field resonant third harmonic surface wave on a half-space of incompressible material of cubic nonlinearity. 
J. Mech. Phys. Solids, 120, 5-15.

\bibitem{Woch08}
Wochner M.S., Hamilton M.F., Ilinskii Y.A., Zabolotskaya E.A. (2008).
Cubic nonlinearity in shear wave beams with different polarizations. 
J. Acoust. Soc. Am., 123, 488-495.

\bibitem{Woch08b}
Wochner, M.S., Hamilton, M.F., Ilinskii, Y.A. and Zabolotskaya, E.A. (2008b). 
Nonlinear torsional wave beams. In AIP Conference Proceedings (Vol. 1022, No. 1, pp. 335-338). AIP.

\bibitem{ZT}
Zabolotskaya, E. A. (1986). Sound beams in a nonlinear isotropic solid. Sov. Phys. Acoust, 32, 296--299.

\bibitem{Z} 
Zabolotskaya, E. A., \& Khokhlov, R. V. (1969). 
Quasi-plane waves in the nonlinear acoustics of confined beams. Sov. Phys. Acoust, 15, 35-40.
 
\bibitem{Zabo04}
Zabolotskaya, E. A., Ilinskii, Y. A., Hamilton, M. F., \& Meegan, G. D. (2004),
Modeling of nonlinear shear waves in soft solids. 
J. Acoust. Soc. Am., 116, 2807-2813.

\end{thebibliography}
\end{document}